\documentclass[12pt]{article}
\usepackage{amscd,amssymb,amsmath,latexsym,enumerate}
\usepackage[mathscr]{euscript}
\usepackage{epsfig}
\usepackage{fancybox}
\usepackage{verbatim}

\usepackage{ esint }

\usepackage{color}
 
\title{Localization and Chern numbers \\
for weakly disordered BdG operators}

\author{Maxim Drabkin, Giuseppe De Nittis, Hermann Schulz-Baldes
\\
\\
{\small Department Mathematik, Universit\"at Erlangen-N\"urnberg, Germany}
}

\date{ }

\newtheorem{theo}{Theorem}

\newtheorem{proposi}{Proposition}
\newtheorem{lemma}{Lemma}
\newtheorem{coro}{Corollary}

\newcommand{\CM}{{\mathbb C}}
\newcommand{\NM}{{\mathbb N}}
\newcommand{\RM}{{\mathbb R}}

\newcommand{\TM}{{\mathbb T}}
\newcommand{\ZM}{{\mathbb Z}}

\newcommand{\EE}{{\bf E}}

\newcommand{\Oo}{{\cal O}}
\newcommand{\Tr}{\mbox{\rm Tr}}

\newcommand{\Nn}{{\cal N}}

\newcommand{\Cc}{{\cal C}}

\newcommand{\Hh}{{\cal H}}

\newcommand{\one}{{\bf 1}}

\newcommand{\redu}{{\mbox{\rm\tiny red}}}
\newcommand{\ph}{{\mbox{\rm\tiny ph}}}

\begin{document}

\maketitle

\begin{abstract}
After a short discussion of various random Bogoliubov-de Gennes (BdG) model operators and the associated physics, the Aizenman-Molchanov method is applied to prove Anderson localization in the weak disorder regime for the spectrum in the central gap. This allows to construct random BdG operators which have localized states in an interval centered at zero energy. Furthermore, techniques for the calculation of Chern numbers are reviewed and applied to two non-trivial BdG operators, the $p+\imath p$ wave and $d+\imath d$ wave superconductors.

\vspace{.1cm}

\hspace{1.3cm} {\it Dedicated to Leonid Pastur on the occasion of his 75th birthday}


\end{abstract}


\section{Introductory comments}
\label{sec-BdGgen}

BdG Hamiltonians describing the electron gas in a superconductor are of the block form
\begin{equation}
\label{eq-BdGfirst}
H_\mu
\;=\;
\frac{1}{2}\;\begin{pmatrix}
h-\mu & \Delta \\ -\overline{\Delta} & -(\overline{h}-\mu)
\end{pmatrix}
\end{equation}
where the operator $h=h^*$ acting on a one-particle complex Hilbert space $\Hh$ with complex structure $\Cc$ describes a single electron, $\mu\in\RM$ is the chemical potential, and $\Delta$, also an operator on $\Hh$, is called the pairing potential or pair creation potential. In \eqref{eq-BdGfirst} and below, the complex conjugate of an operator $A$ on $\Hh$ is defined by $\overline{A}=\Cc A\,\Cc$. The pairing potential satisfies the so-called BdG equation
\begin{equation}
\label{eq-DeltaClassD}
\Delta^*
\;=\;
-\,\overline{\Delta}
\;,
\end{equation}
assuring the self-adjointness of $H_\mu$. Throughout this work both $h$ and $\Delta$ are bounded operators. Hence $H_\mu$ is a bounded self-adjoint operator on the particle-hole Hilbert space $\Hh_{\mbox{\tiny\rm ph}}=\Hh\otimes\CM^2_{\mbox{\tiny\rm ph}}$. The factor $\CM^2_{\mbox{\tiny\rm ph}}$ is called the particle-hole fiber. In the associated grading, the BdG Hamiltonian has the particle hole symmetry (PHS)
\begin{equation}
\label{eq-BdGsymmetry}
K^*\,\overline{H_\mu}\,K
\;=\;
-\,H_\mu
\;,
\qquad
K\;=\;
\begin{pmatrix} 0 & \one \\ \one &  0 \end{pmatrix}
\;.
\end{equation}
The BdG Hamiltonian \eqref{eq-BdGfirst} is obtained from the BCS model by means of a self-consistent mean-field approximation \cite{dG}. In the associated second quantized operator on Fock space (quadratic in the creation and annihilation operators), the off-diagonal entries $\Delta$ and $-\overline{\Delta}$ lead to annihilation and creation of Cooper pairs respectively. Various standard tight-binding models for $\Delta$ are described in Section~\ref{sec-interaction} below. 

\vspace{.2cm}

There are various reasons to consider the operator entries of $H_\mu$ to be random \cite{AZ}. First of all, in a so-called dirty superconductor one can have a random potential just as in any alloy or semiconductor. Moreover, it is reasonable to model the mean field by a random process (even though random in time may seem more adequate). Hence all entries  of \eqref{eq-BdGfirst} can be random operators. For mesoscopic systems it is even reasonable to assume these entries to be random matrices \cite{AZ}. However, in the models considered this paper a spacial structure is conserved by supposing that both $h$ and $\Delta$ only contain finite range hopping operators on $\Hh=\ell^2(\ZM^d)\otimes \CM^r$ where $r$ is the number of internal degrees of freedom and the complex structure is induced by complex conjugation.

\vspace{.2cm}

BdG Hamiltonians having only the symmetry \eqref{eq-BdGsymmetry} are said to be in Class D of the Altland-Zirnbauer (AZ) classification. If furthermore a time-reversal symmetry is imposed, one obtains the Classes AIII and DIII depending on whether spin is even or odd. Particularly interesting are also models with a SU$(2)$ spin rotation invariance. Then \cite{AZ,DS} the Hamiltonian (with odd spin) can be written as a direct sum of spinless Hamiltonians $H_\mu^\redu$ satisfying
\begin{equation}
\label{eq-oddPHS}
I^*\,\overline{H_\mu^\redu}\,I\;=\;-\,H_\mu^\redu
\;,
\qquad
I
\;=\;
\begin{pmatrix}
0 & -\,\one \\ \one & \;\; 0
\end{pmatrix}
\;.
\end{equation}
This is also a PHS, but an odd one because $I^2=-\one$, while the PHS \eqref{eq-BdGsymmetry} is said to be even because $K^2=\one$. Operators with an odd PHS are said to be in the AZ Class C. It is also possible to have a spin rotation symmetry only around one axis (namely, a U$(1)$-symmetry), which can then be combined with a time reversal symmetry (TRS) and this leads to operators lying in other AZ Classes \cite{SRFL}. Let us point out one immediate consequence of the PHS (either even or odd):

\begin{proposi}
\label{prop-specsym}
If $H=H^*$ has a {\rm PHS}, then the spectrum satisfies $\sigma(H)=-\,\sigma(H)$.
\end{proposi}

\noindent Therefore the energy $0$ is a reflection point of the spectrum and hence special. It is furthermore shown in Section~\ref{sec-DOS} below that the integrated density of states for covariant BdG Hamiltonians is symmetric around $0$ and that it is generic that $0$ either lies in a gap or in a pseudo gap. In particular, the situation in \cite{KMM} is non-generic.

\vspace{.2cm}

Since the late 1990's there has been a lot of interest in topological properties of BdG Hamiltonians which can be read of the Fermi projection $P_\mu=\chi(H_\mu\leq 0)$ on particle-hole Hilbert space $\Hh_\ph$, namely the spectral projection on all negative energy states of $H_\mu$. Here the focus is on the two-dimensional case $d=2$. Then the Fermi projection can have non-trivial Chern numbers. For disordered systems, these invariants are defined as in \cite{BES} and enjoy stability properties, see Section~\ref{sec-stability}.  Non-triviality of the Chern number makes the system into a so-called topological insulator \cite{SRFL}. This leads to a number of interesting physical phenomena. For Class D, one has a quantized Wiedemann-Franz law \cite{Vis,SF} and Majorana zero energy states at half-flux tubes \cite{RG}, while for Class C one is in the regime of the spin quantum Hall effect \cite{SMF,RG}. A rigorous analysis of these effects will be provided in \cite{DS}. For periodic models, the Chern numbers can be calculated using the transfer matrices or the Bloch functions. Two techniques to carry out these calculations are discussed and applied in Section~\ref{sec-comp}. 

\vspace{.2cm}

Anderson localization is of importance for topological insulators just as it is for the quantum Hall effect.  It provides localized states near zero energy which are responsible for the stability and measurability of effects related to the topological invariants. For this purpose, it is interesting to have a mathematical proof of Anderson localization in the relevant regimes of adequate models, and to show that the topological invariants are indeed stable. The present paper provides such proofs for BdG models in the weakly disordered regime and shows that localized states near zero energy can be produced by an adequate choice of the parameters, see the discussion after Theorem~\ref{theo-localization} in Section~\ref{sec-localization} resumed in Figure~4. Let us point out that the localization proof  transposes directly to yet other classes of models of interest, the chiral unitary class (AZ Class AIII which also has spectral symmetry as in the one in Proposition~\ref{prop-specsym}) as well as operators with odd time reversal symmetry (AZ Class AII), but no detailed discussion is provided for these cases. For the chiral unitary class this is an important input for the existence of non-commutative higher winding numbers \cite{PS}.

\vspace{.2cm}

On a technical level, this is achieved basically by combining known results. The Aizenman-Molchanov method \cite{AM} in its weak disorder version \cite{Aiz}  provides a framework that can be followed closely, with adequate modifications related to the fact that one has to deal with matrix valued potentials and hopping amplitudes. This has recently be tackled by Elgart, Shamis and Sodin \cite{ESS}, but these authors focused on the strong disorder regime and the models do not seem to cover quite what is needed in connection with the questions addressed above. Furthermore, our arguments seem (to us) a bit more streamlined, and are closer to the original analysis in \cite{Aiz}. The stability of the Chern numbers under disordered perturbation is then obtained just as in \cite{RS}. The multiscale analysis \cite{FS} is another method, historically the first one, to prove localization. This approach has been followed  by Kirsch, M\"uller, Metzger and Gebert in \cite{KMM,GM} for certain models of BdG type in Class CI.

\vspace{.2cm}

\noindent {\bf Acknowledgements.} This work was partially funded by the DFG. G. De Nittis thanks the Humboldt Foundation for financial support. We all thank the UNAM in Cuernavaca for particularly nice office environment while this work was done. 

\section{BdG Hamiltonians in tight-binding representation}
\label{sec-interaction}

This section merely presents the basic tight-binding models used to model and numerically analyze dirty superconductors with particular focus on the form of the pairing potentials. These terms are of relevance both for high-temperature superconductors ({\it e.g.} \cite{WSS,Sca}) as well as for topological insulators \cite{SRFL}. The electron Hilbert space $\Hh$ in \eqref{eq-BdGfirst} is chosen to be $\ell^2(\ZM^d)\otimes\CM^r$. The $r\in\NM$ internal degrees of freedom are used to describe a spin as well as possibly a sublattice degree of freedom, or larger unit cells ({\it e.g.} \cite{ASV} contains a very detailed description of the sublattice degree for the honeycomb lattice as well as various spin-orbit interactions). Let us focus on the two-dimensional case $d=2$, and, just for sake of concreteness, let the one-electron Hamiltonian be given by 
\begin{equation}
\label{eq-elham}
h\;=\;
S_1\,+\,S_1^*\,+\,S_2\,+\,S_2^*\;+\;\lambda\;\sum_{l\in\ZM^2}\,\pi_l^*\, V_l\,\pi_l 
\;.
\end{equation}
Here $S_1$ and $S_2$ are the shift operators on $\ell^2(\ZM^2)$, $\pi_l^*:\CM^r\to \Hh$ is the partial isometry onto the spin and internal degrees of freedom over site $l\in\ZM^d$ and the $V_l=V_l^*$ are i.i.d. matrices of size $r\times r$. To lowest order of approximation, the pairing potential $\Delta$ is often chosen to be translation invariant and the numbers characterizing $\Delta$ are then called the superconducting order parameters. In this section, some examples of such translation invariant pairing potentials are presented. They do not cover all cases studied in the literature \cite{WSS,Sca}, but hopefully the most important ones. Here is the list, expressed in terms of the shift operators and the spin operators $s^1,s^2,s^3$ represented on $\CM^{2s+1}$, which is part of the fiber $\CM^r$. Several of the pairing potentials are graphically represented in Figure 1. Thinking of atomic orbitals, this also explains the nomenclature. 
\begin{align}
& \Delta_s\;=\;
\delta_s\,\imath s^2\;,
&&  s=\tfrac{1}{2}\;\;\mbox{(singlet $s$-wave)}
\\
& \Delta_{s^*}\;=\;
\delta_{s^*}\,(S_1+S_1^*+S_2+S_2^*)\, \imath\, s^2\;,
&&  s=\tfrac{1}{2}\;\;\mbox{(singlet extended $s$-wave)}
\\
& \Delta_{p_x}\;=\;
\delta_{p_x}\,(S_1-S_1^*)\,s^1\;,
&&  s=\tfrac{1}{2}\;\;\mbox{(spinful $p_x$-wave)}
\label{eq-px}
\\
& \Delta_{p\pm\imath p}\;=\;
\delta_{p}\,(S_1-S_1^*\pm\imath(S_2-S_2^*) )\;,
&&  s=0\;\;\mbox{(spinless $p\pm\imath p$-wave)}
\label{eq-pip}
\\
& \Delta_{p}\;=\;\delta_{p}\,s^1\;,
&&  s=\tfrac{1}{2}\;\;\mbox{(spinful $p\pm\imath p$-wave)}
\\
& \Delta'_{p}\;=\;\delta_{p}\,(S_1-S_1^*\pm\imath(S_2-S_2^*)\,s^3 )
\;,
&&  s=\tfrac{1}{2}\;\;\mbox{(triplet $p\pm\imath p$-wave)}
\\
& \Delta_{d_{xy}}\;=\;
\delta_{d_{xy}}\,(S_1-S_1^*)(S_2-S_2^*)\,\imath\, s^2 \;,
&&  s=\tfrac{1}{2}\;\;\mbox{(singlet $d_{xy}$-wave)}
\\
& \Delta_{d_{x^2-y^2}}\;=\;
\delta_{d_{x^2-y^2}}\,(S_1+S_1^* -S_2-S_2^*)\,\imath\,s^2 \;,
&&  s=\tfrac{1}{2}\;\;\mbox{(singlet $d_{x^2-y^2}$-wave)}
\\
& \Delta_{d\pm\imath d}\;=\;\Delta_{d_{x^2-y^2}}\pm\imath \,\Delta_{d_{xy}}
 \;,
&&  s=\tfrac{1}{2}\;\;\mbox{(singlet $d\pm\imath d$-wave)}
\label{eq-did}
\;.
\end{align}
All the constants $\delta$ are real so that one readily checks that \eqref{eq-DeltaClassD} holds in all cases. Here $p$-wave pairing potentials correspond to hopping terms which are anti-symmetric under the change $S_j\leftrightarrow S_j^*$, while $s$-wave and $d$-wave pairing potentials are symmetric under this change. Furthermore, the $s$-wave is rotation symmetric and the $d$-wave odd under a 90 degree rotation $(S_1,S_2,S_1^*,S_2^*)\leftrightarrow (S_2,S_1^*,S_2^*,S_1)$. Next follow two very concrete examples that show that the pairing potential can open a central gap (namely, a gap of $H_\mu$ around zero energy). In Section~\ref{sec-stability} it is discussed under which circumstances these models lead to non-trivial topology of the Bloch bundles.

\begin{figure}
\begin{center}
\includegraphics[height=3.7cm]{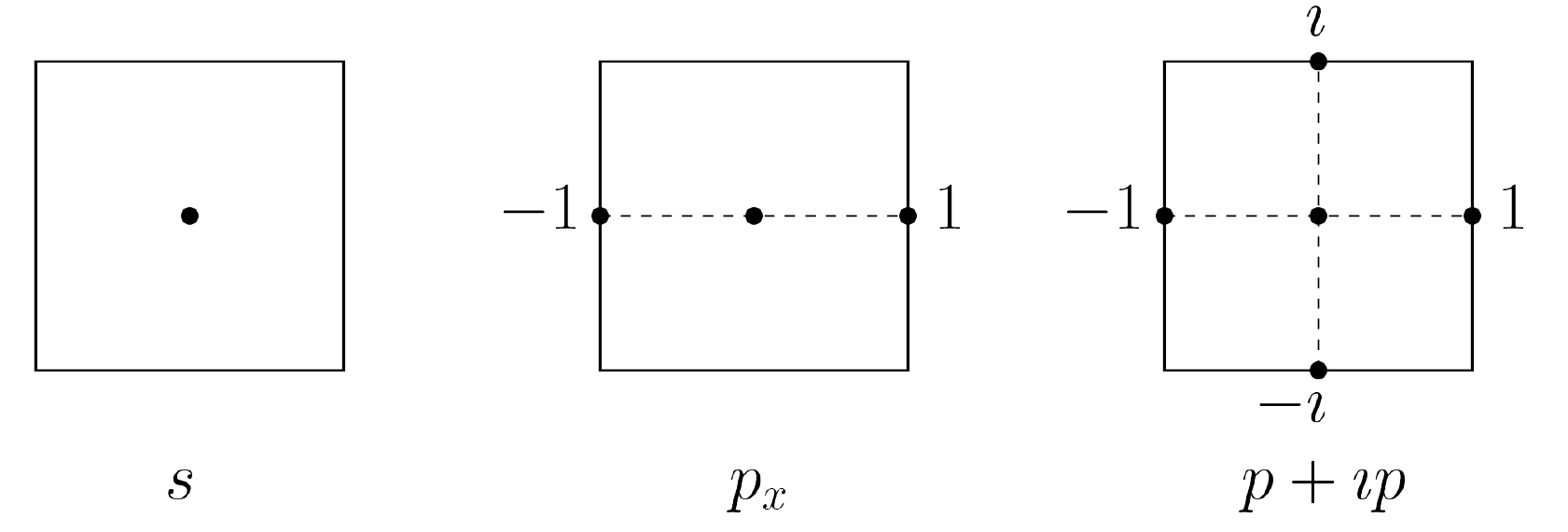}
\vspace{.5cm}
\includegraphics[height=4cm]{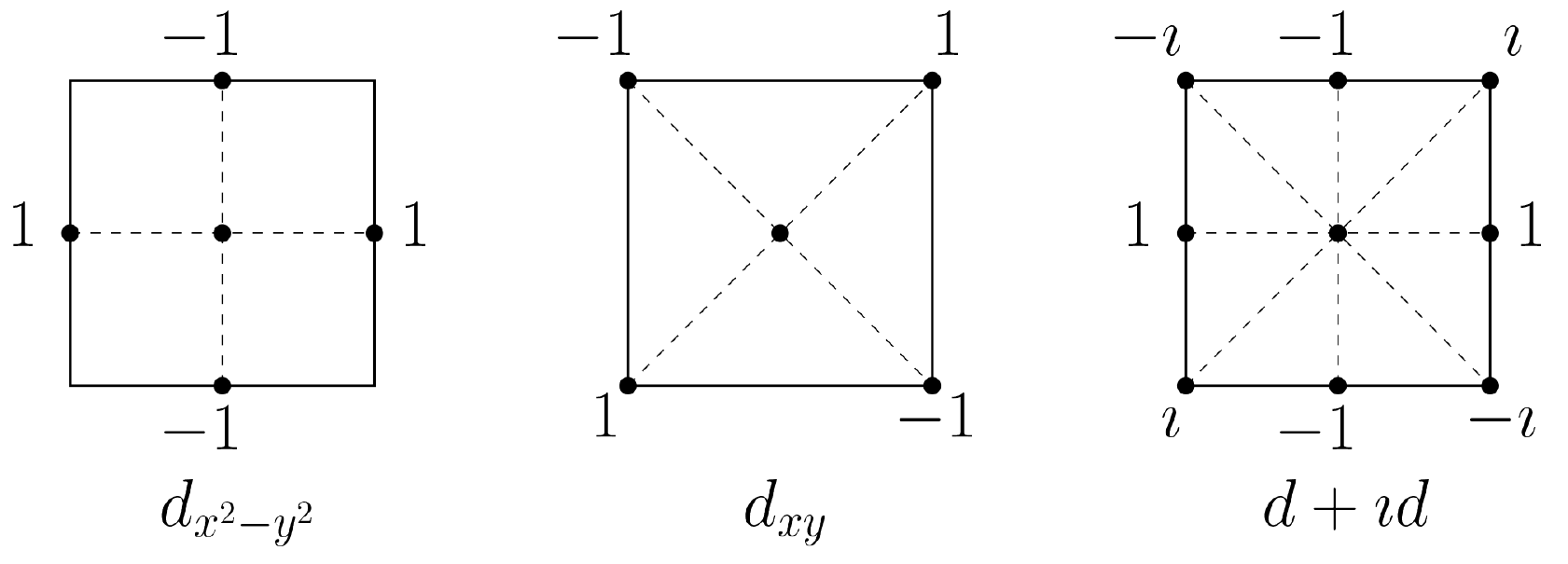}
\caption{\it Graphic representation of various pairing potentials.}
\end{center}
\end{figure}

\vspace{.2cm}

\noindent {\bf Example 1} This example is about a spinless $p\pm\imath p$ model which is hence in Class D and is relevant for the thermal quantum Hall effect (see \cite{Vis,SF,DS} for details). The Hilbert space is simply $\ell^2(\ZM^2)$, namely $r=1$. The one-electron Hamiltonian $h$ is \eqref{eq-elham} with $\lambda=0$ and the pairing potential given by \eqref{eq-pip}. Thus the BdG Hamiltonian is 
$$
H_\mu
\;=\;
\frac{1}{2}\;
\begin{pmatrix}
S_1+S_1^*+S_2+S_2^* -\mu & \delta_{p}\,(S_1-S_1^*\pm\imath(S_2-S_2^*) )
\\
\delta_{p}\,(S_1^*-S_1\pm\imath(S_2-S_2^*) ) &
-S_1-S_1^*-S_2-S_2^*+\mu
\end{pmatrix}
\;.
$$
After discrete Fourier transform
$$
H_\mu(k)
\;=\;
\begin{pmatrix}
\cos(k_1)+\cos(k_2) -\frac{\mu}{2} & \delta_{p}\,(\imath\sin(k_1)\mp \sin(k_2))
\\
\delta_{p}\,(-\imath\sin(k_1)\mp \sin(k_2))
 &
-\cos(k_1)-\cos(k_2)+\frac{\mu}{2}
\end{pmatrix}
\;.
$$
Hence the eigenvalues are
$$
E_\eta(k)
\;=\;
\eta\,\left(
(\cos(k_1)+\cos(k_2) -\tfrac{\mu}{2})^2
+
\delta_{p}^2\,(\sin^2(k_1)+ \sin^2(k_2))
\right)^{\frac{1}{2}}
\;,
\qquad
\eta\in\{-1,1\}
\;.
$$
This shows that a central gap of size $g(\delta_p,\mu)=2\,\min_{k\in\TM^2}\,E_+(k)$ opens  if $|\mu|\not=0,4$ and $\delta_p\not=0$. If $\mu=0$, the gap is closed for all $\delta_p$. Furthermore, the gap satisfies the upper bound $g(\delta_p,\mu)\leq 2\,E_+(0,\pi)=|\mu|$. For $\mu$ sufficiently small compared to $\delta_p$, one even has $g(\delta_p,\mu)=|\mu|$ so that the gap closes linearly in $|\mu|$.
\hfill $\diamond$

\vspace{.2cm}

\noindent {\bf Example 2}
A $d\pm\imath d$ model is of interest in connection with the spin quantum Hall effect (see \cite{SMF,RG,DS}). Here the spin is $s=\frac{1}{2}$ so that $r=2$. Again $h$ is the discrete Laplacian (tensorized with $\one$ on the spin degree of freedom) and $\Delta$ is given by \eqref{eq-did}. As this contains an $\imath s^2$ in the spin component, the interaction as well as $h$ are $\mbox{\rm SU}(2)$ invariant. For sake of simplicity let us assume $\delta_{d_{x^2-y^2}}=\delta_{d_{x,y}}=\delta_d$.  The BdG Hamiltonian $H_\mu$ is a $4\times 4$ matrix, but it can be written as a direct sum $H_\mu=H_\mu^+\oplus H_\mu^-$ of $2\times 2$ operator matrices $H_\mu^\pm$ satisfying the odd PHS $I^*\overline{H_\mu^\pm}I=-H_\mu^\pm$, so that $H_\mu^\pm$ are in Class C. Their Fourier transforms are given by
$$
H_\mu^\pm(k)
\,=\,
\begin{pmatrix}
\cos(k_1)+\cos(k_2) -\frac{\mu}{2} & \!\!\!\! \delta_{d}\,(\cos(k_1)-\cos(k_2)\mp \imath\sin(k_1)\sin(k_2))
\\
\delta_{p}\,(\cos(k_1)-\cos(k_2)\pm \imath\sin(k_1)\sin(k_2))
 &
-\cos(k_1)-\cos(k_2)+\frac{\mu}{2}
\end{pmatrix}
\,.
$$
Such a direct sum decomposition is always possible in presence of an $\mbox{\rm SU}(2)$-invariance \cite{AZ,DS}. The two Bloch bands of $H_\mu^\pm$, indexed by $\eta\in\{-1,1\}$, are 
\begin{equation}
\label{eq:EnBand}
E_\eta(k)
\;=\;
\eta\,\left(
(\cos(k_1)+\cos(k_2) -\tfrac{\mu}{2})^2
+
\delta_{d}^2\,(\cos(k_1)\cos(k_2)-1)^2
\right)^{\frac{1}{2}}
\;.
\end{equation}
Again a central gap opens for $\delta_d\not= 0$ and $|\mu|\neq 0,4$ and its size satisfies $g(\delta_d,\mu)\leq 2 \,E_+(0,0)$ and $g(\delta_d,\mu)\leq 2 \,E_+(\pi,\pi)$, so that $g(\delta_d,\mu)\leq  |4-|\mu||$.
\hfill $\diamond$

\section{Density of states of covariant BdG Hamiltonians}
\label{sec-DOS}

\begin{figure}
\begin{center}
\includegraphics[height=7cm]{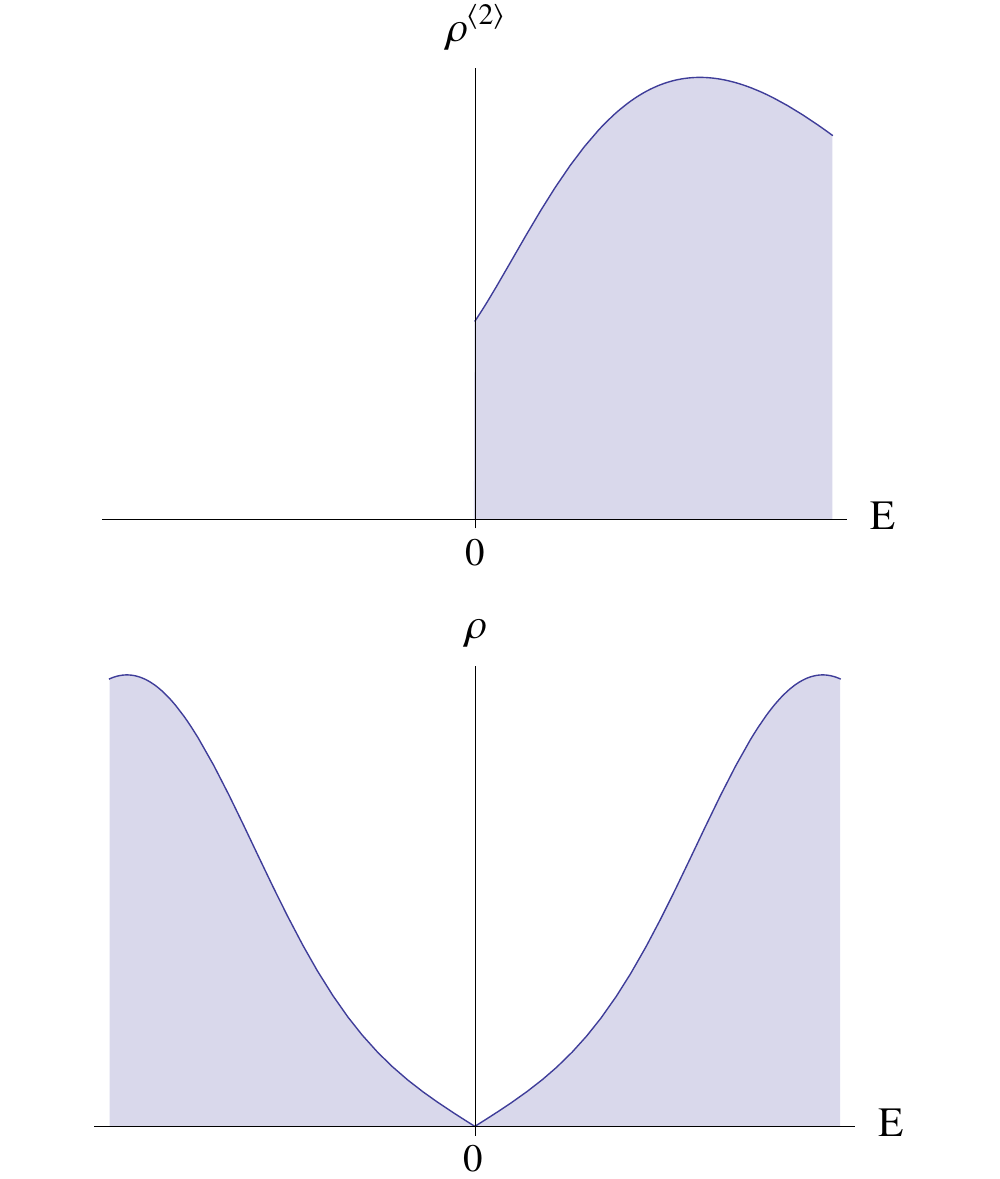}
\hspace{-.1cm}
\includegraphics[height=7cm]{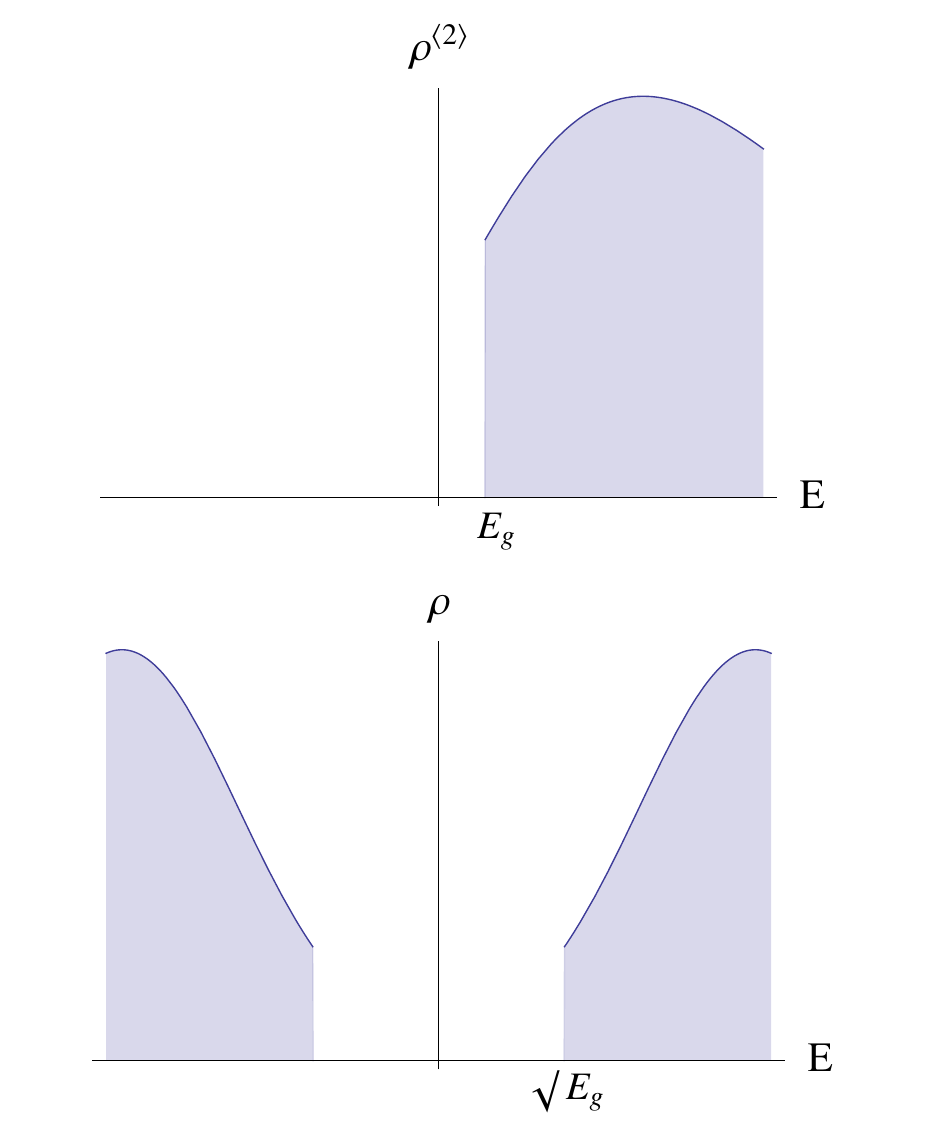}
\caption{\it Schematic representation of the {\rm DOS} $\rho^{(2)}$ and corresponding $\rho$ in the two generic situations for periodic operators in dimension $d=2$: with pseudo gap and true gap.}
\end{center}
\end{figure}

This short section discusses an extension of Proposition~\ref{prop-specsym}, namely a symmetry of the integrated density of states (IDS) of a covariant family of BdG operators (with even or odd PHS). Let $S_j$, $j=1,\ldots,d$, denote the shifts on $\Hh=\ell^2(\ZM^d)\otimes\CM^r$, naturally extended to the particle-hole Hilbert space $\Hh_{\mbox{\tiny\rm ph}}=\Hh\otimes\CM^2_{\mbox{\tiny\rm ph}}$. A strongly continuous family $A=(A_{\omega})_{\omega\in\Omega}$ of bounded operators on $\Hh_{\mbox{\rm\tiny ph}}$ is called covariant if
\begin{equation}
\label{eq-covariance}
S_j
A(\omega)
S_j^{-1} 
\;=\;
A(T_j \omega)
\mbox{ , }
\qquad j=1,\ldots,d
\mbox{ . }
\end{equation}
Here $\Omega$ is a compact space (of disorder or crystaline configurations) which is furnished with an action $T=(T_1,\ldots,T_d)$ of the translation group $\ZM^d$. Furthermore, there is given an invariant and ergodic probability measure on $\Omega$. Let us now consider a family $H=(H(\omega))_{\omega\in\Omega}$ of BdG Hamiltonians satisfying the covariance relation \eqref{eq-covariance}. By general principles, this implies that $H$ has a well-defined IDS $\Nn$. Usually, one chooses the normalization condition $\Nn(-\infty)=0$, but here we rather choose to impose $\Nn(0)=0$ in view of Proposition~\ref{prop-specsym}. With this normalization one has for $E\geq 0$
$$
\Nn(E)
\;=\;\EE\;\Tr\;\langle 0|\chi_{[0,E]}(H(\omega))|0\rangle
\;,
\qquad
\Nn(-E)
\;=\;- \,\EE\;\Tr \;\langle 0|\chi_{[-E,0]}(H(\omega))|0\rangle
\;,
$$
where $\Tr$ denotes the trace over $\CM^r\otimes\CM_{\mbox{\rm\tiny ph}}^2$ and $\EE$ the average over the invariant measure on $\Omega$, and the notation $\langle n|A|m\rangle=\pi_n A\pi_m^*\in\mbox{Mat}(r\times r,\CM)$ is used in order to stress the similarity with the scalar case $r=1$. Now the BdG symmetry implies
\begin{equation}
\label{eq-IDSsym}
\Nn(E)\;=\;-\,\Nn(-E)\;.
\end{equation}
%
%
%
Furthermore, the IDS $\Nn$ can be nicely expressed in terms of the IDS $\Nn^{(2)}$ of the positive operator $H^2$ given by
$$
\Nn^{(2)}(E)\;=\;
\EE\;\Tr\;\langle 0|\chi_{[0,E]}(H(\omega)^2)|0\rangle
\;.
$$
Using the symmetry \eqref{eq-IDSsym} one finds for $E\geq 0$
\begin{equation}
\label{eq-IDSsym2}
\Nn(E)\;=\;\frac{1}{2}\;\Nn^{(2)}(E^2)
\;.
\end{equation}
If $\Nn$ and $\Nn^{(2)}$ are absolutely continuous with density of states (DOS) $\rho(E)$ and $\rho^{(2)}(E)$ respectively, then one deduces
$$
\rho(E)
\;=\;
|E|\;\rho^{(2)}(E^2)
\;.
$$
This shows that generically a periodic BdG operator in dimension $d=2$ either has a gap in the DOS or a so-called linear pseudo-gap, namely the DOS $\rho$ vanishes linearly at $0$, up to lower order corrections. These two generic cases are illustrated in Figure~2, while non-generic cases are given in Figure~3.

\begin{figure}
\begin{center}
\includegraphics[height=7cm]{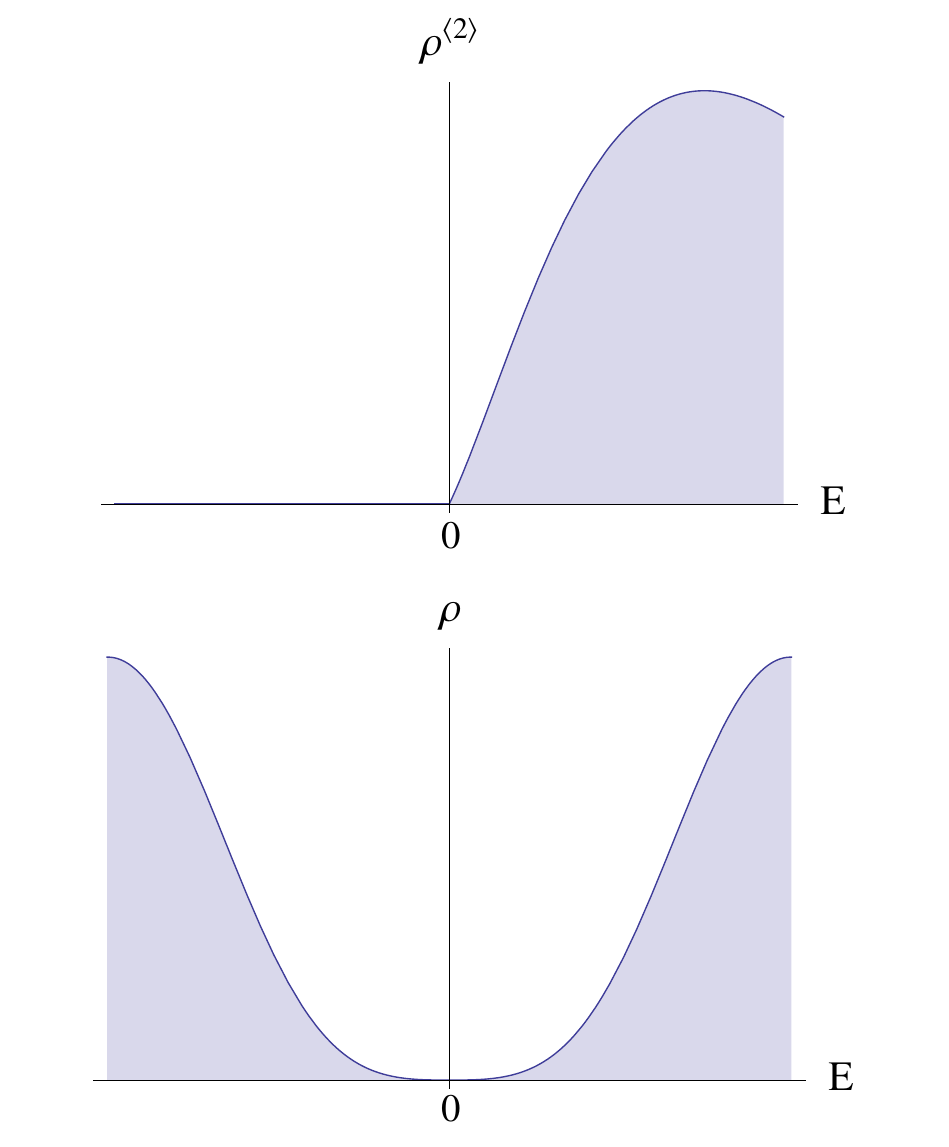}
\hspace{-.5cm}
\includegraphics[height=7cm]{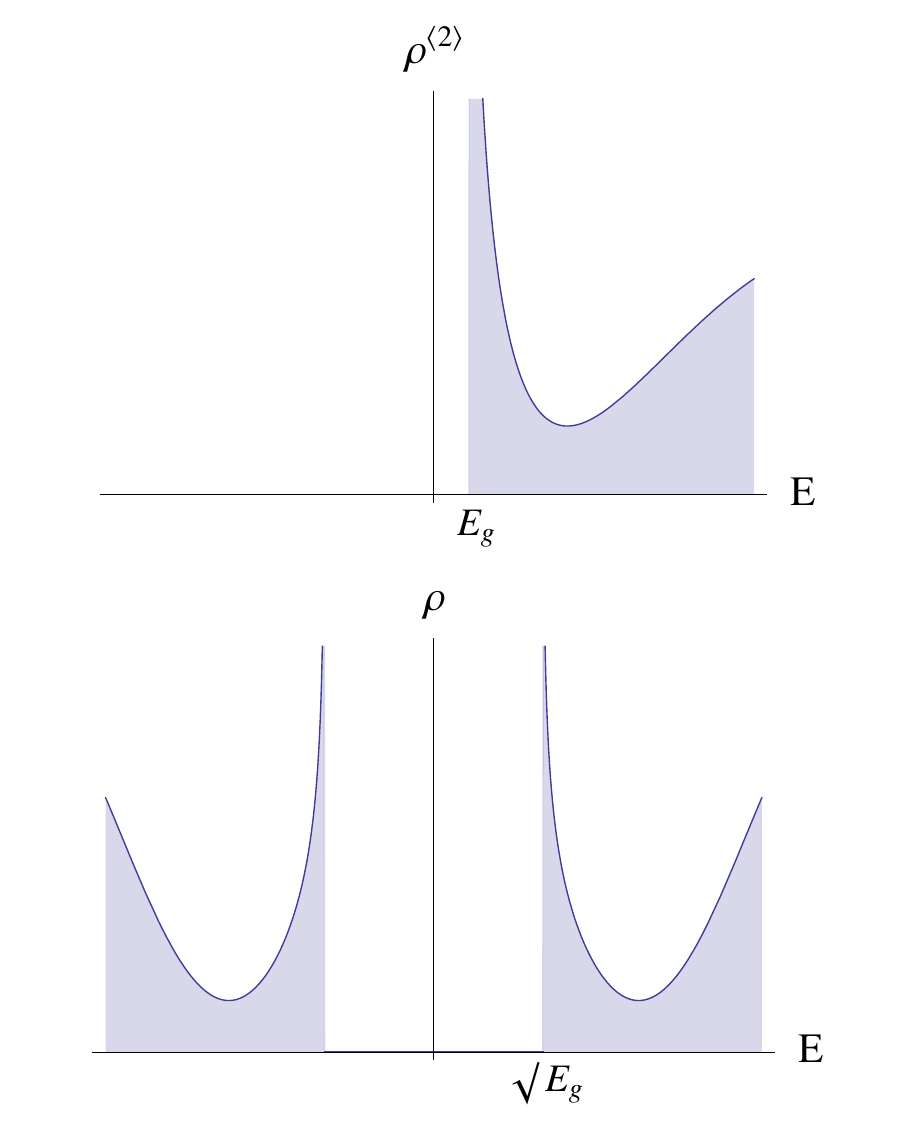}
\hspace{-.5cm}
\includegraphics[height=7cm]{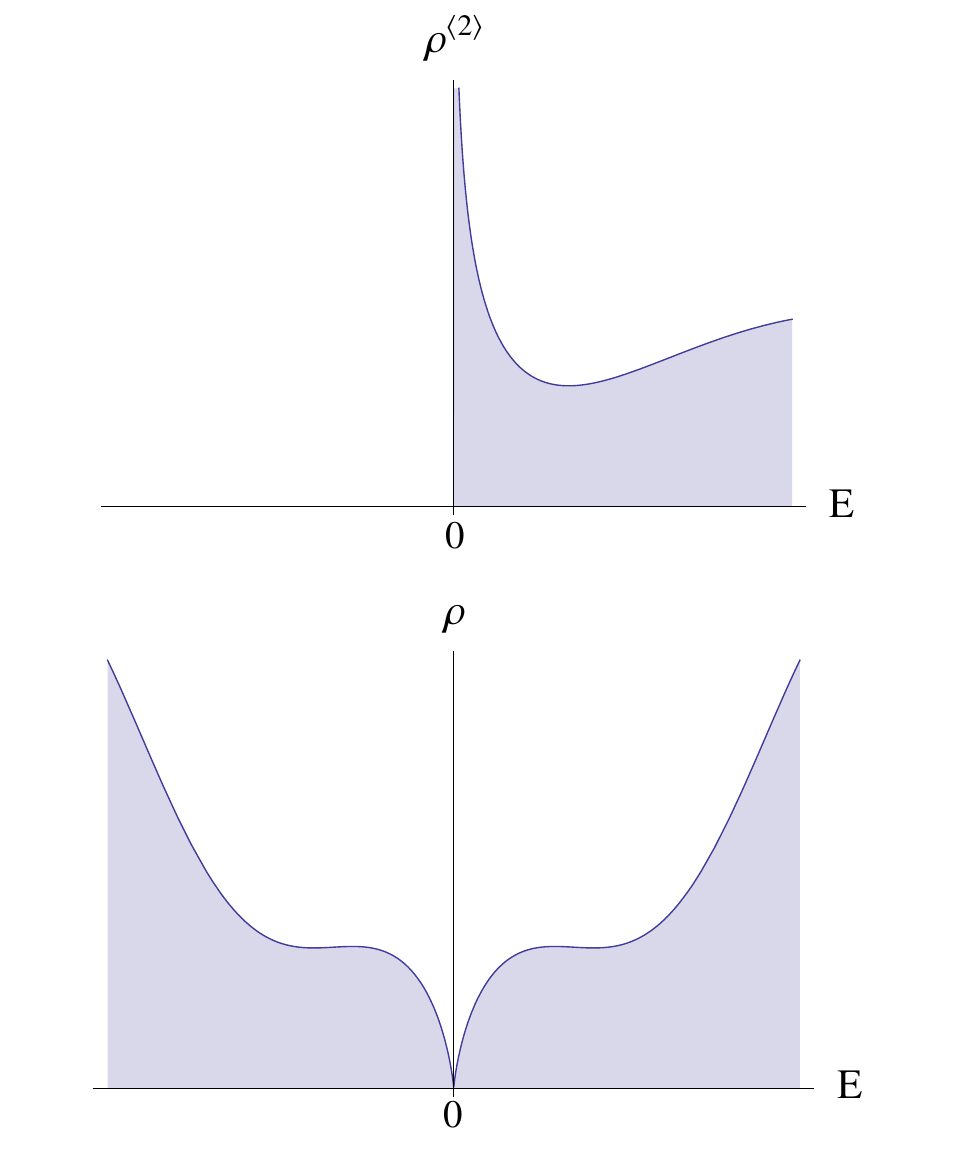}
\caption{\it Schematic representation of the {\rm DOS} $\rho^{(2)}$ and corresponding $\rho$ in various non-generic cases for periodic operators in dimension $d=2$. Note that the first picture would be generic in dimension $d=4$ {\rm (}linear vanishing of the {\rm DOS} at the bottom of the spectrum of $H^2${\rm )}, while the third one is generic in dimension $d=1$ {\rm (}square root singularity at the bottom of the spectrum of $H^2${\rm )}.}
\end{center}
\end{figure}

\section{Localization for BdG Hamiltonians}
\label{sec-localization}

The one-particle BdG operators considered in this section act on the particle-hole Hilbert space $\Hh_{\mbox{\tiny\rm ph}}=\Hh\otimes\CM^2_{\mbox{\tiny\rm ph}}$ with $\Hh=\ell^2(\ZM^d)\otimes\CM^r$. It is of the form
\begin{equation}
\label{eq-HamAnd}
H_{\mu,\lambda}
\;=\;
H_{\mu,0}\,+\,\lambda\,V\;,
\qquad
V\;=\;\sum_{l\in\ZM^d}\,\sum_{|j|\leq R}\,v_{j,l}\,
\pi_{l+j}^*\,W_j\,\pi_{l}
\end{equation}
where $\lambda>0$ is a coupling constant, $\pi_l:\Hh_{\mbox{\tiny\rm ph}}\to \CM^r\otimes \CM^2_{\mbox{\tiny\rm ph}}$ is the partial isometry onto the spin and particle-hole space over site $l\in\ZM^d$, and for each $j$ the matrices $W_j$ act on $\CM^{2r}=\CM^r\otimes \CM^2_{\mbox{\tiny\rm ph}}$, and the $v_{j,l}$ are real random numbers. It is assumed that $W_{-j}=W_j^*$ and $v_{j,l}=v_{-j,l+j}$ which assures that $V$ is self-adjoint. The operator $H_{\mu,\lambda}$ is supposed to satisfy the even or odd PHS and possibly TRS, depending on which symmetry class is to be described. Furthermore, $H_{\mu,0}$  is supposed to be a translation invariant (or at least periodic) BdG operator with chemical potential $\mu$ and interest will be mainly in the situation where $H_{\mu,0}$ already contains a central spectral gap, {\it e.g.} opened by a constant pairing potential. The framework of \eqref{eq-HamAnd} allows, in particular, to cover all the models discussed in Section~\ref{sec-interaction}, but also allows for random or periodic pairing potentials and spin orbit interactions. Some examples of matrices $W_j$  are
$$
W_{(0,0)}\;=\;
\begin{pmatrix}
\one & 0 \\ 0 & -\one
\end{pmatrix}
\;,
\qquad
W_{(1,0)}
\;=\;
\begin{pmatrix}
0 & \one \\ -\one & 0 
\end{pmatrix}
\;,
\qquad
W_{(0,1)}
\;=\;
\begin{pmatrix}
0 & \imath\,\one \\ \imath\,\one & 0 
\end{pmatrix}
\;.
$$
Then $W_{(0,0)}$ allows to model a random potential contained in $h$ as in \eqref{eq-elham}, but also allows to vary the chemical potential via $H_{\mu',\lambda}=H_{\mu,\lambda}+(\mu'-\mu)\;\one_{\ell^2(\ZM^d)}\otimes W_{(0,0)}$. On the other hand, $W_{(1,0)}$ and $W_{(0,1)}$ can be used to describe a random pairing potential for spinless $p\pm\imath p$ waves. Other random pairing potentials form the list of Section~\ref{sec-interaction} can also be described by operators of the form \eqref{eq-HamAnd}.  There is a number of technical hypothesis imposed on all these objects:

\vspace{.2cm}

\noindent {\bf Hypothesis:} {\it The $v_{j,l}=v_{-j,l+j}$ are independent random variables for all $l\in\ZM^d$ and $|j|\leq R$. For each fixed $j$ the distributions $\nu_j$ of the $v_{j,l}$ are identically {\rm (}in $l${\rm )} and uniformly $\alpha$-H\"older continuous with $1-\nu_j([-a,a])$ decaying in $a$ faster than any polynomial.}

\vspace{.2cm}

Let us introduce the $2r\times 2r$ Green matrices at energy $z\in\CM$ by
$$
G^z_{\mu,\lambda}(n,m)
\;=\;
\pi_n \,(z-H_{\mu,\lambda})^{-1}\,\pi_m^*
\;.
$$

\begin{theo}
\label{theo-localization}
Suppose that the random {\rm BdG} Hamiltonian satisfies the above hypothesis. Let $I\subset \RM$ be a compact spectral interval with $D=\mbox{\rm dist}(I,\sigma(H_{\mu,0}))>0$ and let $s<1$ be sufficiently small. Then there exist constants $C_1$, $C_2$ and $C_3$ such that for $\lambda\leq C_1D^{1+\frac{1}{s}}$  the Hilbert-Schmidt norm of the Green matrix satisfies 
\begin{equation}
\label{eq-AMbound}
\EE\bigl(\|G^z_{\mu,\lambda}(n,m)\|_2^s\bigr)
\;\leq\;
C_2\;e^{-s\,C_3\,D\, |n-m|}
\;,
\end{equation}
for all $z=E+\imath\epsilon$ with $E\in I$ and independently of $\epsilon\not=0$.
\end{theo}

Before starting with the proof let us investigate under which circumstance the statement of the theorem is not void, namely when there are energies in the spectrum of $H_{\mu,\lambda}$ satisfying the hypothesis needed to prove the exponential decay (outside of the spectrum it already holds due to a simple Combes-Thomas estimate stated in Proposition~\ref{prop-CT} below). This point was already discussed by Aizenman \cite{Aiz}, but here it is, moreover, of particular importance to produce models with localized states at zero energy. Suppose that one only adds a random potential (that is, only the term with $W_{(0,0)}$ above). If the support of $\nu=\nu_{(0,0)}$ is $[-r,r]$, then the almost sure spectrum of $H_{\mu,\lambda}$ grows from the band edges at least as $\lambda r$ until the central gap closes. More precisely, by a standard probabilistic argument the spectrum of the BdG operator with the above random potential is equal to the spectrum of the deterministic BdG model with chemical potential shifted from $\mu>0$ to $\mu-\lambda r$. Assuming that the gap is closed for the periodic model with vanishing chemical potential (as in the two examples), the gap of the random model is closed for $\lambda\geq \frac{\mu}{r}$. Furthermore, it closes linearly in the two examples considered above (and thus also Fig.~4), but this is not important for the following. On the other hand, the condition $\lambda\leq C_1D^{1+\frac{1}{s}}$ assures a localization regime by Theorem~\ref{theo-localization} and this condition is independent of $r$ (as long as the moments of the distribution of the random potential remain uniformly bounded). Hence choosing $r$ sufficiently large guarantees the existence of an interval centered at zero energy containing only localized states. Let us mention that this does not address the important question about the fate of the states near a pseudo gap of a periodic model when a random perturbation is added (raising the density of states in the pseudo gap of the periodic model). Indeed it was supposed here that the gap for the model without disorder is opened by some mechanism such as the pairing potential.

\begin{figure}
\begin{center}
\includegraphics[height=6cm]{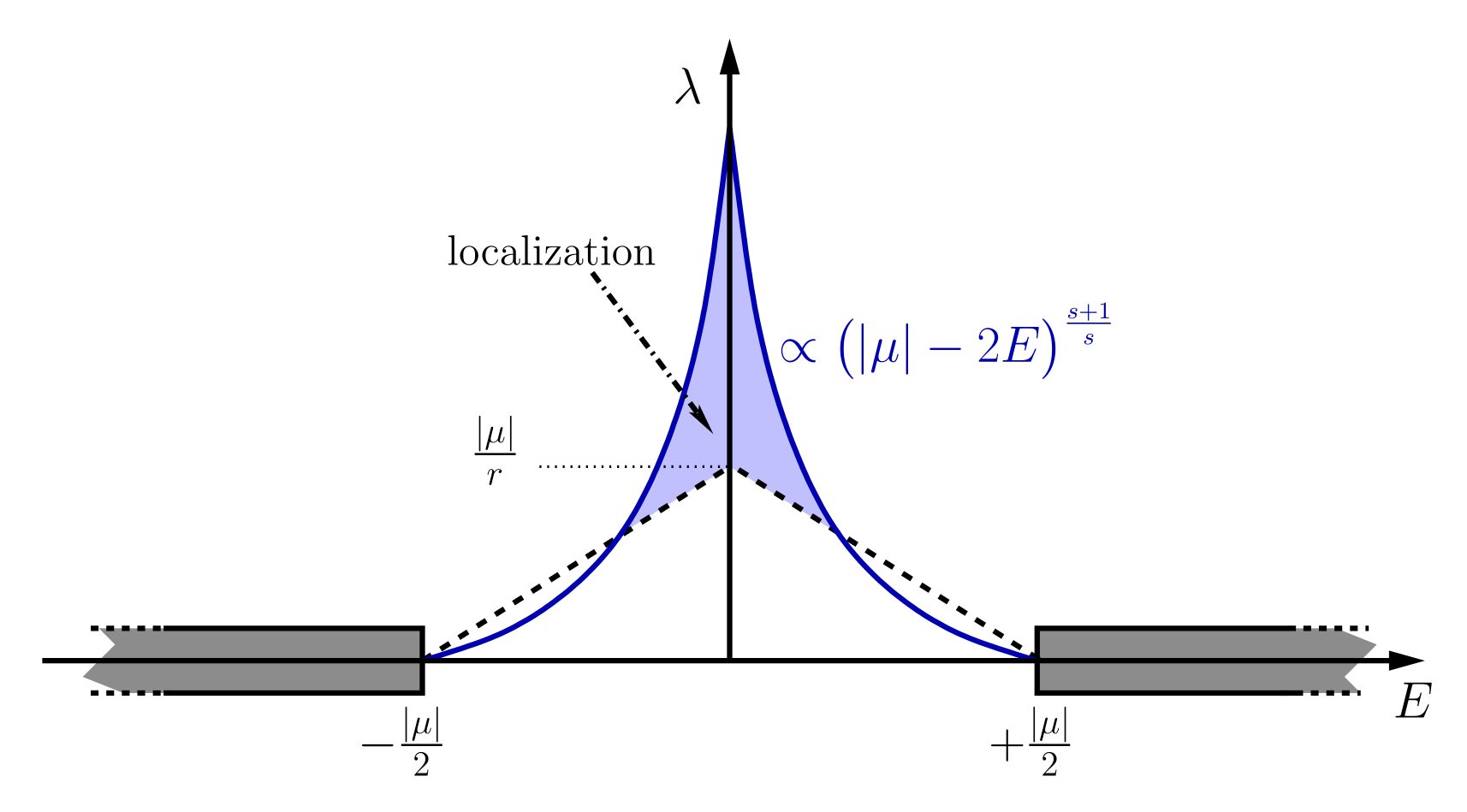}
\caption{\it Spectrum and localization regime attained by {\rm Theorem~\ref{theo-localization}} for the $p+\imath p$ wave superconductor model with diagonal potential disorder and $|\mu|\ll \delta_p$, plotted in an energy-disorder coupling plane.}
\end{center}
\end{figure}

\vspace{.2cm}

Another comment is that the Hypothesis does not contain any minimal coupling condition. In fact, this would be necessary in a strong coupling regime, but for weak disorder results one only shows that the decay of the free Green functions (given by Combes-Thomas) is conserved under adequate perturbations. Before proceeding to the proof, let us note a consequence of \eqref{eq-AMbound} that is important for the definition of the Chern numbers in the next section. The following corollary can be deduced by a technique put forward in \cite{AG}, see also Theorem~5.1 in \cite{PS} for a detailed argument. Physically interesting is only the case $E=0$ and for this case it pends on Theorem~\ref{theo-localization} only if there is spectrum at $0$.

\begin{coro}
\label{coro-localization}
Let $I\subset\RM$ be a compact energy interval on which {\rm \eqref{eq-AMbound}} holds. Then
for any $E\in I$ which is not a boundary point of $I$ and any $\alpha>0$, the Fermi projection $P_\mu(E)=\chi(H_\mu\leq E)$ satisfies
\begin{equation}
\label{eq-Pdecay}
\EE\;\|\,\langle n| P_\mu(E)|m\rangle\,\|\;<\;C_4\; |n-m|^{-\alpha}
\;,
\end{equation}
where $C_4$ is a constant depending on $\alpha$. 
\end{coro}

\vspace{.2cm}

The proof of Theorem~\ref{theo-localization} uses the resolvent identity 
$$
G^z_{\mu,\lambda}(n,m)
\;=\;
G^z_{\mu,0}(n,m)
\,+\,
\lambda\,
\sum_{l\in\ZM^d}\,\sum_{|j|\leq R}\,v_{j,l}\,G^z_{\mu,0}(n,l+j)\,W_{j}\,G^z_{\mu,\lambda}(j,m)
\;.
$$
These matrices will be estimated using the Hilbert-Schmidt norm $\|\,.\,\|_2$ for matrices acting on the fibers $\CM^{2r}$:
$$
\|G^z_{\mu,\lambda}(n,m)\|_2
\;\leq\;
\|G^z_{\mu,0}(n,m)\|_2
\,+\,
\lambda\,
\sum_{l\in\ZM^d}\,\sum_{|j|\leq R}\,|v_{j,l}|\,\|G^z_{\mu,0}(n,l+j)\|_2\,\|W_{j}\|_2\,
\| G^z_{\mu,\lambda}(l,m)\|_2
\;.
$$
Now let us take the expectation value over the randomness, as well as the $s$th power, where $s<1$. Using $(\sum_l|a_l|)^s\leq\sum_l|a_l|^s$ one obtains
\begin{align*}
& \EE\bigl(\|G^z_{\mu,\lambda}(n,m)\|_2^s\bigr)
\\
& \;\;\;\;\leq\;
\|G^z_{\mu,0}(n,m)\|_2^s
\,+\,
\lambda^s\,
\sum_{l\in\ZM^d}\,\sum_{|j|\leq R}\,\,\|G^z_{\mu,0}(n,l+j)\|_2^s\;\|W_{j}\|_2^s\;
\EE\bigl(|v_{j,l}|^s\,\| G^z_{\mu,\lambda}(l,m)\|_2^s\bigr)
\;.
\end{align*}
The aim is now to prove exponential decay in $|n-m|$ of the quantity
$$
\tau^z_\lambda(n,m)
\;=\;
\EE\bigl(\|G^z_{\mu,\lambda}(n,m)\|_2^s\bigr)
\;,
$$
for adequate energies $z=E+\imath 0$. The basic input is the exponential decay of $\tau^z_0(n,m)$ which is obtained by a so-called Combes-Thomas estimate (which will be used only for $H_{\mu,0}$ below).

\begin{proposi}
\label{prop-CT}
Suppose that $H_{\mu,\lambda}$ has finite range $R$, namely $\pi_n^*H_{\mu,\lambda}\pi_m=0$ for $|n-m|>R$. Then there are constants $C_5$, $C_6$ and $C_7$ such that
$$
\|G^z_{\mu,\lambda}(n,m)\|_2
\;\leq\;
\frac{C_5}{D(z)}\,\exp\Bigl(-C_6\mbox{\rm arcsinh}(C_7D(z))\,|n-m|\Bigr)
\;,
\qquad
D(z)=\mbox{\rm dist}(z,\sigma(H_{\mu,\lambda}))
\;.
$$
\end{proposi}

\noindent {\bf Proof.} Let us drop the indices $\mu,\lambda$ and choose one direction $j\in\{1,\ldots,d\}$. For $\eta\in\RM$, set $H(\eta)=e^{\eta X_j}He^{-\eta X_j}$ where $X_j$ is the $j$th component of the position operator on $\ell^2(\ZM^d)$, naturally extended to $\Hh_{\mbox{\tiny\rm ph}}$. Then one has the following norm estimate in terms of the matrices  $H_{n,m}=\pi_n^*H\pi_m$
\begin{eqnarray*}
\|\,H(\eta)-H\,\|^2
& = &
\sup_{\|\phi\|=1}
\sum_{n,m}
\left\|H_{n,m}(1-e^{(n_j-m_j)\eta})\phi_m\right\|^2
\\
& \leq &
\sup_{\|\phi\|=1}
\sum_{m}
\left(\sum_n\,\left\|H_{n,m}\right\|^2\;|1-e^{(n_j-m_j)\eta}|^2\right)
\,\|\phi_m\|^2
\\
& \leq &
\sup_{m}
\left(\sum_n\,\left\|H_{n,m}\right\|^2\;|1-e^{(n_j-m_j)\eta}|^2\right)
\;.
\end{eqnarray*}
Now a uniform upper bound on the $H_{n,m}$ and $|1-e^x|\leq 2\sinh(|x|)$ implies
$$
\|\,H(\eta)-H\,\|
\;\leq\;
C_8\,\sinh(R|\eta|)
\;.
$$
Now recall the bound $\|(\one+B)^{-1}\|\leq 1+\|B(\one+B)^{-1}\|$ holding for any  operator $B$ for any operator with invertible $\one+B$.  Using this for $B=(H(\eta)-H)(H-z)^{-1}$, one has
$$
\left\|(H(\eta)-z)^{-1}\right\|
\;\leq\;
\left(\,\left\|(H-z)^{-1}\right\|^{-1}\,-\,\|H(\eta)-H\|\,\right)^{-1}.
$$
Since $\|(H-z)^{-1}\|\leq D(z)^{-1}$, the choice $R|\eta|=\mbox{\rm arcsinh}\left(D(z)/2\,C_7\right)$ therefore leads to the bound $\|(H(\eta)-z)^{-1}\|\leq 2/D(z)$. The desired estimate now follows from the identity $\langle n|(H-z)^{-1}|m\rangle =e^{\eta(m_j-n_j)}\,\langle n|(H(\eta)-z)^{-1}|m\rangle$ by choosing the adequate sign for $\eta$.
\hfill $\Box$

\vspace{.2cm}

In order to use the Combes-Thomas estimate as a starting point for a perturbative analysis, one has to prove the following decorrelation estimate
\begin{equation}
\label{eq-decorr}
\EE\bigl(|v_{j,l}|^s\,\| G^z_{\mu,\lambda}(l,m)\|_2^s\bigr)
\;\leq\;
C_{\alpha,s}\;
\EE\bigl(|v_{j,l}|^s\bigr)\,\EE\bigl(\| G^z_{\mu,\lambda}(l,m)\|_2^s\bigr)
\;.
\end{equation}
Its proof is deferred to the end of this paragraph. Once this is proved, the following  subharmonicity argument applied to $\tau(n)=\tau^z_\lambda(n,m)$ and $\tau_0(n)=\tau^z_0(n,m)$ for every fixed $m$ concludes the proof.

\begin{lemma}
\label{lem-subharm}
Suppose that $\tau_0:\ZM^d\to \RM_{\geq 0}$ satisfies
$$
\tau_0(n)
\;\leq\;c_1\,e^{-c_2|n|}\;.
$$
Then, if another function $\tau:\ZM^d\to \RM_{\geq 0}$ satisfies the subharmonicity estimate
$$
\tau(n)
\;\leq\;
\tau_0(n)
\;+\;
c_3\,\sum_{l\in\ZM^d}\,
\tau_0(n-l)\tau(l)
$$
with $c_3\leq c_4c_2/c_1$ and $c_4$ only depending on the dimension $d$, this function satisfies
$$
\tau(n)
\;\leq\;c_5\,e^{-c_2|n|}\;,
$$
with $c_5$ of order $1$.
\end{lemma}

\noindent {\bf Proof.} Let us introduce an operator $T:\ell^\infty(\ZM^d)\to\ell^\infty(\ZM^d)$ by
$$
(T\tau)(n)
\;=\;
c_3\,\sum_{l\in\ZM^d}\,
\tau_0(n-l)\tau(l)
\;.
$$
Then
$$
\|T\|_{\infty\to\infty}
\;=\;
\sum_{n\in\ZM^d} c_3\,\tau_0(n)
\;\leq\;
c_1\,c_3\,\sum_{n\in\ZM^d}\,e^{-c_2|n|}
\;<\;1
\;,
$$
where the last equality holds for $c_3c_1/c_2$ sufficiently small. Iterative application of the subharmonicity inequality then shows
$$
\tau(n)
\;\leq\;
\sum_{i=0}^\infty\;(T^i\tau_0)(n)
\;.
$$
Let us set $b(n)=e^{-c_2|n|}$ and consider this as a multiplication operator on as well as an element of $\ell^\infty(\ZM^d)$. Telescoping then shows
$$
\tau(n)
\;\leq\;
b(n)\;\sum_{i=0}^\infty\;\Bigl((b^{-1}T\,b)^i\,\frac{\tau_0}{b}\Bigr)(n)
\;\leq\;
b(n)\;\sum_{i=0}^\infty\;\|b^{-1}T\,b\|_{\infty\to\infty}^i\,\Big\|\frac{\tau_0}{b}\Big\|_\infty
\;.
$$
Now $\frac{\tau_0}{b}$ is bounded by hypothesis, and
$$
\|b^{-1}T\,b\|_{\infty\to\infty}
\;\leq\;
\sup_n\sum_l e^{c_2|n|}\,c_3\,e^{-c_2|n-l|}\,e^{-c_2|l|}
\;\leq\;
\frac{c_3\,c_6}{c_2}
\;.
$$
Now supposing again that $c_3$ is sufficiently small such that $c_3c_6<c_2$, the result follows by summing the geometric series in the above estimate.
\hfill $\Box$

\vspace{.2cm}

Let us briefly show how this allows to conclude the proof of Theorem~\ref{theo-localization}. Due to Proposition~\ref{prop-CT}, one can choose $c_1=(C_5/D)^s$ and $c_2=sC_6C_7D C_9$ with $C_9$ depending on the size of $I$. Furthermore, $c_3=\lambda^s C_{\alpha,s} C_{10}$ with $C_{10}$ bounding the moments of the random variable and the norms of $W_j$. Then the smallness assumption in Lemma~\ref{lem-subharm} reads  $\lambda^s C_{\alpha,s} C_{10}\leq c_4sC_6C_7D C_9 (D/C_5)^s$, which is the small coupling assumption in Theorem~\ref{theo-localization}.

\vspace{.2cm}

For the proof of the decorrelation estimate \eqref{eq-decorr} the dependence of $G^z_{\mu,\lambda}(l,m)$ on $v_{j,l}$ has to be determined in an explicit manner. This can be done via a standard perturbative formula which is written in a manner that can be immediately applied to the present model if one sets $v=v_{j,l}$ and $W=W_j$.

\begin{lemma}
\label{lem-Krein}
Let us consider the following splitting of the Hamiltonian
$$
H_{\mu,\lambda}
\;=\;
\widetilde{H}_{\mu,\lambda}\,+\,\lambda\,v\,(W+W^*)
\;,
$$
with an operator $\widetilde{H}_{\mu,\lambda}$ which does not depend on the real parameter $v$.  Let $\pi_v:\Hh\to\mbox{\rm Ran}(W+W^*)$ be the partial isometry onto $\mbox{\rm Ran}(W+W^*)$. Then for any $z\in\CM\setminus\RM$
$$
\bigl(z-H_{\mu,\lambda}\bigr)^{-1}
\;=\;
\bigl(z-\widetilde{H}_{\mu,\lambda}\bigr)^{-1}
\,-\,
\bigl(z-\widetilde{H}_{\mu,\lambda}\bigr)^{-1}
\,\pi_v^*
\,T_v\,\pi_v\,
\bigl(z-\widetilde{H}_{\mu,\lambda}\bigr)^{-1}
\;,
$$
where the finite-dimensional $T$-matrix is given by
$$
T_v
\;=\;
\left(
\bigl(\lambda \,v\,\pi_v(W+W^*)\pi_v^*\bigr)^{-1}
-
\pi_v\bigl(z-\widetilde{H}_{\mu,\lambda}\bigr)^{-1}\pi_v^*
\right)^{-1}
\;.
$$
\end{lemma}

\vspace{.2cm}

The standard algebraic proof of Lemma~\ref{lem-Krein} can be found {\it e.g.} in \cite[Lemma~8]{BS}. Let us point out that $W+W^*=\pi_v^*\pi_v(W+W^*)\pi_v^*\pi_v$ and that $\pi_v(W+W^*)\pi_v^*$ is invertible by construction. This inverse is again a self-adjoint operator. As the operator $\pi_v\bigl(z-\widetilde{H}_{\mu,\lambda}\bigr)^{-1}\pi_v^*$ has positive imaginary part, the inverse defining $T_v$ actually exists. Next let us suppose $W$ and $W^*$ have finite range. Then $T_v$ is a finite dimensional matrix which can be calculated using Lagrange formula. This shows that $T_v$ is a rational function of $v$. Applying this to the situation sketched above, it follows that $G^z_{\mu,\lambda}(l,m)$ is a rational function of $v_{j,l}$. Therefore the decorrelation estimate \eqref{eq-decorr} follows from the following lemma.

\vspace{.2cm}

\begin{proposi}
\label{prop-decop}
Let $f$ and $g$ be rational functions given by a fraction of polynomials with degree at most $N$.  If $\nu$ is a uniformly $\alpha$-H\"older continuous measure with $\nu([-R,R]^c)$  decaying faster than any polynomial, and $s<\alpha/2N$, then uniformly in $f$ and $g$ for some constant $C_{\alpha,s}$ depending only on $s$ and $\alpha$ 
$$
\EE_\nu(|fg|^s)\;\leq\;C_{\alpha,s}\,\EE_\nu(|f|^s)\;\EE_\nu(|g|^s)
\;,
$$
\end{proposi}

\noindent {\bf Proof.} 
First of all, it is possible to approximate $\nu$ with a measure with support contained in $[-R,R]$ because $|fg|^s$, $|f|^s$ and $|g|^s$ grow at most polynomially and have integral singularities  so that all factors can be arbitrarily well approximated for large $R$. Thus from now on the support of $\nu$ lies in $[-R,R]$. Let us start from the Cauchy-Schwarz inequality 
$$
\EE_\nu(|fg|^s)
\;\leq\;
\EE_\nu(|f|^{2s})^{\frac{1}{2}}\;\EE_\nu(|g|^{2s})^{\frac{1}{2}}
\;.
$$
Hence it is sufficient to show that
$$
\frac{\EE_\nu(|f|^{2s})^{\frac{1}{2}}}{\EE_\nu(|f|^s)}\;\leq\;(C_{\nu,s})^{\frac{1}{2}}
\;,
$$
for all rational functions $f$. Let $a=(a_1,\ldots,a_{N'})$ and $b=(b_1,\ldots, b_{N''})$ be the zeros of the numerator and denominator respectively with $N',N''\leq N$, namely
$$
f(x)
\;=\;
\frac{\prod_{n'=1}^{N'}(x-a_{n'})}{\prod_{n=1}^{N''}(x-b_{n})}
\;.
$$
Let us introduce the function
$$
F(a,b)
\;=\;
\frac{\EE_\nu(|f|^{2s})^{\frac{1}{2}}}{\EE_\nu(|f|^s)}
\;.
$$
It will be shown that $F$ is continuous in $(a,b)$ and is bounded by a uniform constant outside of the ball $|(a,b)|_\infty\leq 2R$. The continuity of $(a,b)\mapsto \EE_\nu(|f|^s)$ and $(a,b)\mapsto \EE_\nu(|f|^{2s})$ follows from the integrability of the singularities (resulting from the H\"older continuity of $\nu$ with sufficiently large $\alpha$) and the fact that the integral of a continuous family of integrable functions is again continuous. Furthermore, the function $(a,b)\mapsto \EE_\nu(|f|^s)$ is bounded from below by $0$ (uniformly on every compact). Both of these facts require some standard analytical verification using the H\"older continuity of $\nu$. Combined they show the continuity of $F$. It thus follows that $F$ is bounded on every compact set, in particular on $|(a,b)|_\infty\leq 2R$. 

\vspace{.2cm}

Now suppose that $|(a,b)|_\infty>2R$. Let $a_1,\ldots,a_{M'}$ and $b_1,\ldots,b_{M''}$ be the zeros with modulus smaller than or equal to $2R$ (after having renumbered). Let us set
$$
g(x)
\;=\;
\frac{\prod_{n'=1}^{M'}(x-a_{n'})}{\prod_{n=1}^{M''}(x-b_{n})}
\;,
\qquad
G(a,b)\;=\;\frac{\EE_\nu(|g|^{2s})^{\frac{1}{2}}}{\EE_\nu(|g|^s)}
\;.
$$
Note that $G$ only depends on the first $M'$ and $M''$ of the $a$'s and $b$'s. By the same argument as for $F$,  $G$ is uniformly bounded on the set defined by $|a_{n'}|\leq 2R$ for $n'=1,\ldots,M'$ and $|b_n|\leq 2R$ for $n=1,\ldots,M''$. For $n'>M'$ and $n>M''$ one has $|a_{n'}|\geq 2R$ and $|b_n|\geq 2R$ and therefore, for every $x\in [-R,R]$, the following bounds hold 
$$
|a_{n'}|-R\;\leq\;|a_{n'}-x|\;\leq\;|a_{n'}|+R\;,
\qquad
|b_{n}|-R\;\leq\;|b_{n}-x|\;\leq\;|b_{n}|+R\;.
$$
Consequently
$$
F(a,b)
\;\leq\;
\left(
\prod_{n'=M'+1}^{N'}
\frac{|a_{n'}|+R}{|a_{n'}|-R}\;
\prod_{n=M''+1}^{N''}
\frac{|b_{n}|+R}{|b_{n}|-R}
\right)^s
\;G(a,b)
\;\leq\;3^{(N'+N'')s}\;G(a,b)
\;.
$$
As there are a finite number of possibilities to choose $G$ and the bound is independent of $R$, the claim follows.
\hfill $\Box$


\section{Chern numbers and their stability}
\label{sec-stability}

First let us recall \cite{BES} the definition of the Chern number of a covariant family $P=(P(\omega))_{\omega\in\Omega}$ of projections on $\ell^2(\ZM^2)\otimes \CM^r$:
\begin{equation}
\label{eq-Chern}
\mbox{\rm Ch}(P)
\;=\;
2\pi\imath\;\EE\;\Tr\;\langle 0|P[[X_2, P],[X_1, P]]|0\rangle
\;,
\end{equation}
where $X_1$ and $X_2$ are the two components of the position operators on $\ell^2(\ZM^2)$, $\EE$ denotes the disorder average, and the projection is required to satisfy the so-called Sobolev condition
\begin{equation}
\label{eq-Sobolev}
\sum_{j=1,2}\;\EE\;\Tr\;\langle 0||[X_j, P]|^2|0\rangle\;<\;\infty
\;.
\end{equation}
This condition assures that \eqref{eq-Chern} is well-defined. For the (particle-hole space) Fermi projection $P_{\mu,\lambda}=\chi(H_{\mu,\lambda}\leq 0)$ the condition \eqref{eq-Sobolev} holds if the central gap remains open (by the Helffer-Sj\"ostrand formula combined with a Combes-Thomas estimate) or, due to Corollary~\ref{coro-localization}, if $E=0$ lies in the Aizenman-Molchanov localization regime of $H_{\mu,\lambda}$. On the other hand, the condition \eqref{eq-Sobolev} also assures that $\mbox{\rm Ch}(P)$ is an integer given by the index of a Fredholm operator \cite{BES}. Hence one expects $\mbox{\rm Ch}(P)$ to have some homotopy invariance properties. In particular, one may expect $\mbox{\rm Ch}(P_{\mu,\lambda})$ to be independent of the disorder coupling constant $\lambda$ and the chemical potential $\mu$ under adequate conditions. For quantum Hall systems, this was checked in \cite{RS} and we claim here that the argument directly carries over to the BdG case to prove the following.

\begin{theo}
\label{theo-homotopy}
Let ${\cal R}\subset \{(\mu,\lambda)\in\RM^2\}$ be a connected set such that for every $\lambda$ the energy $E=0$ lies in an open interval for which the bound {\rm \eqref{eq-AMbound}} holds for the {\rm BdG} Hamiltonian $H_{\mu,\lambda}$ of the form {\rm \eqref{eq-HamAnd}}.  Then $(\mu,\lambda)\in {\cal R}\mapsto \mbox{\rm Ch}(P_{\mu,\lambda})\in\ZM$ is constant.
\end{theo}

Due to this stability result, it is particularly important to calculate the Chern number without disorder, that is, for a periodic system with a central gap. This is the object of the next section.


\section{Computation of Chern numbers}
\label{sec-comp}

Two methods for the calculation are briefly presented in this section and applied to the two examples of Section~\ref{sec-interaction}. The first one from \cite{ASV} applies to periodic BdG Hamiltonians $H_\mu$ containing only nearest neighbor and next nearest neighbor hopping terms but arbitrarily (large) fibers. It uses merely the transfer matrices combined with basic numerics. The second, more conventional method applies whenever it is possible to write the Hamiltinian as linear combination of Clifford algebra generators \cite{DL}. In order to deal with the examples one merely has to use Pauli matrices.
As to the first method, one begins with a partial discrete Fourier transform, say in the $1$-direction. Rewriting the fibers of the Hamiltonian as
$$
H_\mu(k_1)
\;=\;
{S}_2^*\,a(k_1)\,+\,b(k_1)\,+a(k_1)^*\,{S}_2
\;,
$$
defines $2r\times 2r$ matrices $a(k_1)$ and $b(k_1)$. Whenever $a(k_1)$ is invertible (which is almost surely the case), one next sets
$$
T(k_1)
\;=\;
\begin{pmatrix}
-b(k_1)a(k_1)^{-1} & -a(k_1)^* \\ a(k_1)^{-1} & 0
\end{pmatrix}
\;.
$$
This is a $4r\times 4r$ matrix which conserves $I$ given in \eqref{eq-oddPHS}, namely $T(k_1)^*IT(k_1)=I$. The generalized eigenspaces of $T(k_1)$ associated with all eigenvalues of modulus strictly less than $1$ (namely the contracting ones) constitute an $2r$-dimensional $I$-Lagrangian plane in $\CM^{4r}$. Let a basis of this space form the column vectors of a $4r\times 2r$ matrix $\Phi(k_1)$ and then define a $2r\times 2r$ matrix $U(k_1)$ by
\begin{equation}
\label{eq-UPhiCalc}
U(k_1)\;=\;\binom{\one}{\imath\,\one}^*\Phi(k_1)\left( \binom{\one}{-\imath\,\one}^*\Phi(k_1)\right)^{-1}
\;.
\end{equation}
It turns out that $U(k_1)$ is unitary and that the following holds.

\begin{theo}
\label{theo-Cherncalc} {\rm \cite{ASV}} Let $0$ lie in a gap of $H_\mu$ and set $P_\mu=\chi(H_\mu\leq 0)$. Then
$$
\mbox{\rm Ch}(P_\mu)
\;=\;
\int^{\pi}_{-\pi}\frac{dk_1}{2\pi\imath}\;\Tr
\left(
U(k_1)^*\partial_{k_1}U(k_1)
\right)
\;.
$$
\end{theo}

Let us apply this theorem to calculate the Chern number of the $p+\imath p$ model discussed in Example~1 in Section~\ref{sec-interaction}. Then $r=1$ so the unitary of size $2\times 2$ that is then calculated numerically from the contracting eigenvector of the $4\times 4$ matrix $T(k_1)$ as a function of $k_1\in [-\pi,\pi]$. The phase of its eigenvalues is plotted for three sets of parameters in Figure~5. 


\begin{figure}
\begin{center}
\includegraphics[height=5cm]{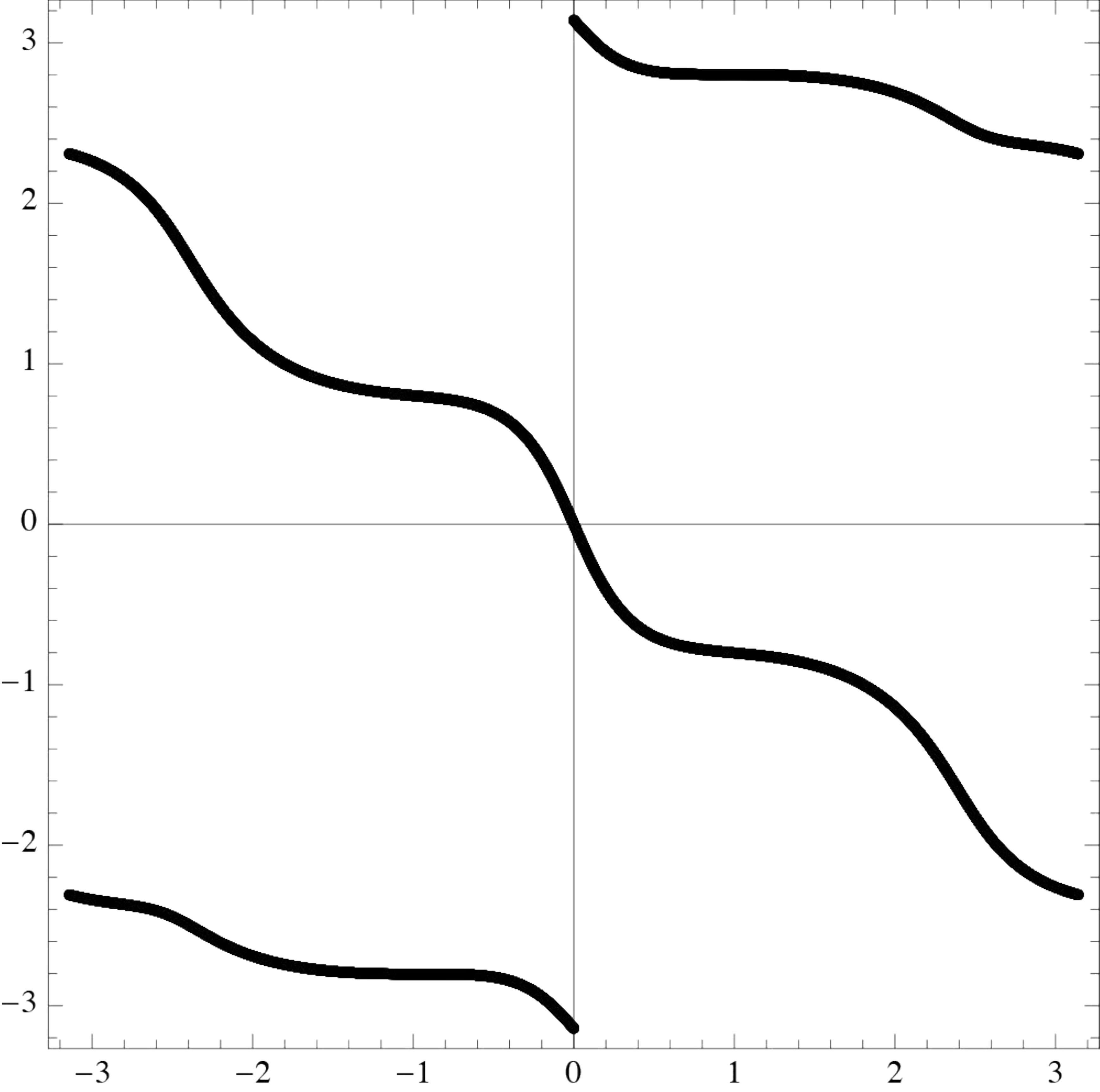}
\hspace{.4cm}
\includegraphics[height=5cm]{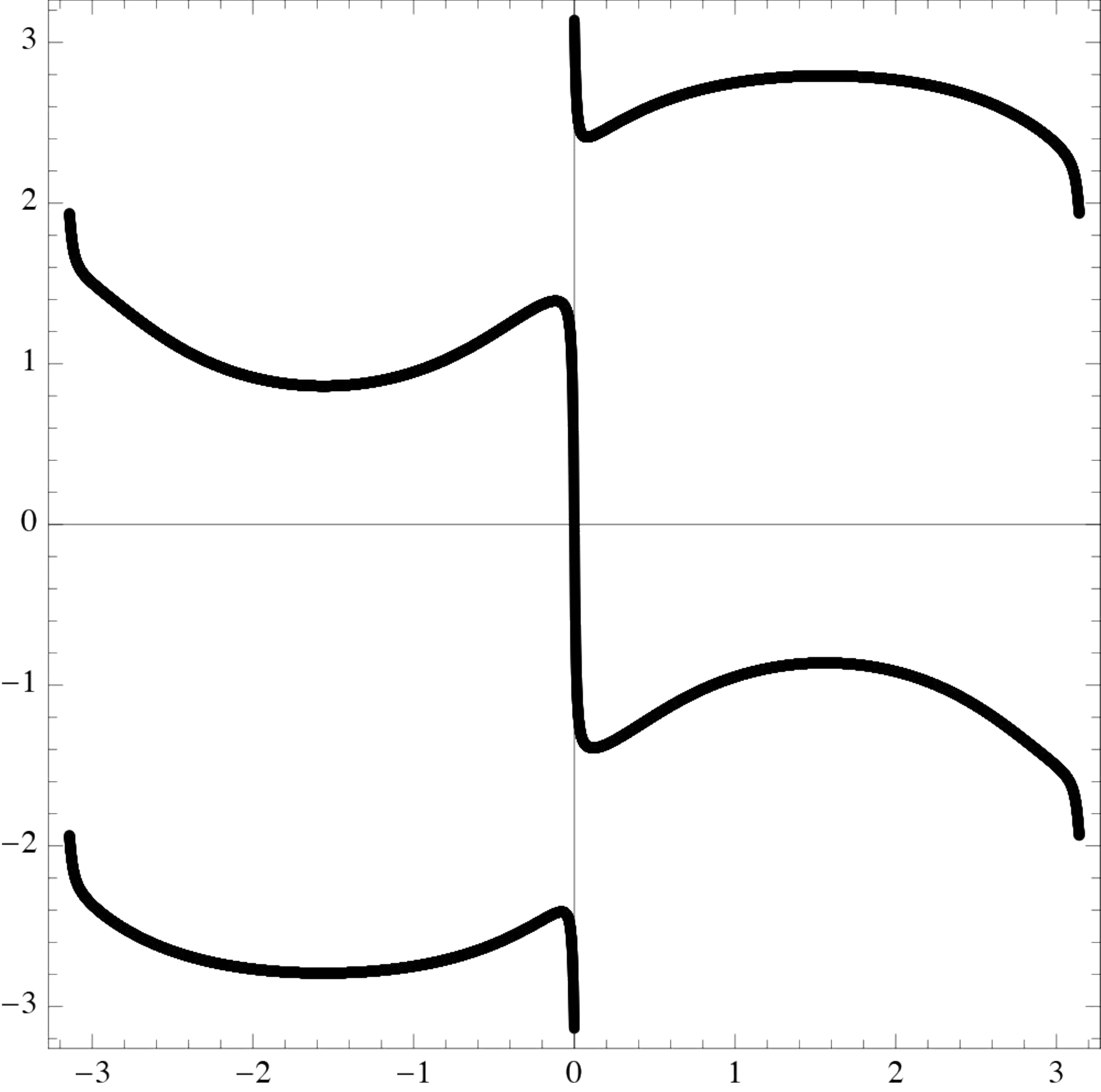}
\hspace{.4cm}
\includegraphics[height=5cm]{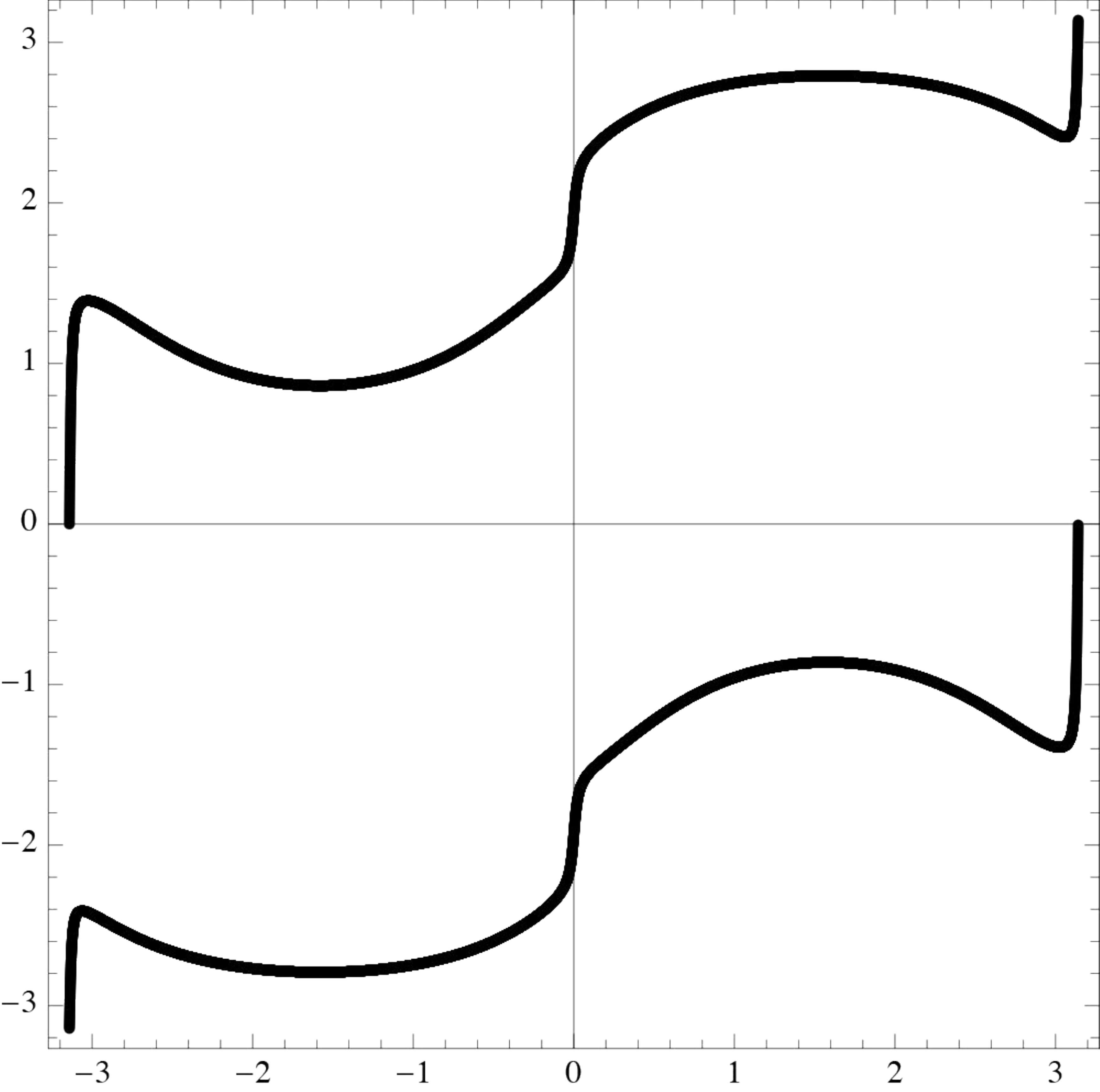}
\caption{\it Plot of the phase $k_1\in[-\pi,\pi]\mapsto U(k_1)$ for the clean $p+\imath p$ wave superconductor described in {\rm Example~1}, with $(\delta_p,\mu)=(0.3,-0.5)$, $(\delta_p,\mu)=(0.3,-0.01)$ and $(\delta_p,\mu)=(0.3,0.01)$. The Chern numbers of $\mbox{\rm Ch}(P_\mu)$ are $-1$, $-1$ and $1$ respectively. Note that at $\mu=0$ the central gap closes and becomes a pseudo gap, so that there is a transition at this value.}
\end{center}
\end{figure}


\vspace{.2cm}

The second method is illustrated by calculating the Chern number of the Fermi projection of the $d+\imath d$-wave Hamiltonians $H^\pm_\mu$ given in Example~2 of Section~\ref{sec-interaction}. First rewrite the Hamiltonian as a linear combination of the Pauli matrices
$$
H^\pm_\mu(k)\;=\;
\begin{pmatrix}
p_3(k) & p_1(k)\mp\imath\,p_2(k)
\\
p_1(k)\pm\imath\,p_2(k) & -p_3(k)
\end{pmatrix}
\;,
$$
where
$$
p_1(k)
\;=\;
\delta_d(\cos(k_1)-\cos(k_2))
\;,
\quad
p_2(k)
\;=\;
\delta_d\sin(k_1)\sin(k_2)
\;,
\quad
p_3(k)
\;=\;
\cos(k_1)+\cos(k_2)-\tfrac{\mu}{2}
\;.
$$
The lower Bloch band is $E_-(k)=-(p_1(k)^2+p_2(k)^2+p_3(k)^2)^{\frac{1}{2}}$ and a central gap opens for $\delta_d\not= 0$ and $|\mu|\neq 0,4$, as already pointed out above. The normalized eigenfunction for the lower eigenvalue, spanning the range of the Fermi projection, can be written in two ways
$$
\psi(k)
\;=\;
C_-(k)\;
\begin{pmatrix}
p_1(k)\mp\imath p_2(k)
\\
E_-(k)-p_3(k)
\end{pmatrix}
\;,
\qquad
\phi(k)
\;=\;
C_+(k)\;
\begin{pmatrix}
E_-(k)+p_3(k)
\\
p_1(k)\pm\imath p_2(k)
\\
\end{pmatrix}
\;.
$$
where $C_\pm(k)=(2E_-(k)(E_-(k)\pm p_3(k)))^{-\frac{1}{2}}$. However, both of these vector functions may vanish for certain values of $k$ (at which then the normalization constants are singular). In fact, $p_1(k)=p_2(k)=0$ holds at $k_I=(0,0)$ and $k_{II}=(\pi,\pi)$. For $\mu>4$, one has $E_-(k)+p_3(k)<0$ so that $\phi$ defines a global section, implying that the bundle is trivializable and has vanishing Chern number. Similarly, for $\mu<-4$, $\psi$ is a global section so that again the Chern number vanishes. Now let us come to the case $|\mu|<4$ where both $\phi$ and $\psi$ have zeros in $k_I$ and $k_{II}$ respectively. Let us introduce the transition function $\theta$ defined on $\TM^2/\{k_I,k_{II}\}$ by
$$
\phi(k)\;=\;
e^{\imath \theta(k)}\;\psi(k)
\;,
\qquad
\theta(k)\;=\;\arctan\left(\frac{\pm\, p_2(k)}{p_1(k)}\right)
\;.
$$
Now set $P_\mu^\pm=\chi(H_\mu^\pm<0)$. By \cite[Sect. 20]{BT} and using the closed path $\Gamma$ given by $t\in[0,2\pi)\mapsto\big(\epsilon\cos (t),\epsilon\sin (t)\big)$ with small $\epsilon>0$,
$$
{\rm Ch}(P_\mu^\pm)
\;=\;
\frac{1}{2\pi}\varointctrclockwise_{\Gamma}{\rm d}\theta
\;=\;
\frac{\epsilon}{2\pi}\int_{0}^{2\pi}{\rm d}t\;\Big(-\,\Theta^\epsilon_1(t)\sin( t)\;+\;\Theta^\epsilon_2(t)\cos (t)\Big)
\;,
$$
where $\Theta^\epsilon_j(t)=\big(\partial_{k_j}\theta\big)(\epsilon\cos (t),\epsilon\sin (t))$.  A straightforward computation   provides
$$
\Theta^\epsilon_1(t)\;=\;\pm\;\frac{2\sin (t)}{\epsilon}\;+\; \Oo(\epsilon)\;,
\qquad 
\Theta^\epsilon_2(t)\;=\;\mp\,\frac{2\cos (t)}{\epsilon}\;+\; \Oo(\epsilon)
\;.
$$
Replacing and taking the limit $\epsilon\to 0$ shows
$$
{\rm Ch}(P_\mu^\pm)\;=\;\mp\; 2\;,
\;
\qquad\text{if}\quad |\mu|<4\;.
$$

\vspace{-.6cm}


\end{document}